# Fracture of dual crosslink gels with permanent and transient crosslinks

Koichi Mayumi[a,b**], Jingyi Guo[c], Tetsuharu Narita[a,b], Chung Yuen Hui[c], and Costantino Creton[a,b*]

[a]Sciences et Ingénierie de la Matière Molle, CNRS UMR 7615, École Supérieure de Physique et de Chimie Industrielles de la Ville de Paris (ESPCI), ParisTech, PSL Research University, 10 rue Vauquelin, F-75231 Paris cedex 05, France

[b]SIMM, UPMC Univ Paris 06, Sorbonne-Universités, 10 rue Vauquelin, F-75231 Paris cedex 05, France

[c]Department of Mechanical and Aerospace Engineering, Cornell University, Ithaca NY 14853, USA.

* Corresponding author

** Current Address: Department of Advanced Materials Science, Graduate School of Frontier Sciences, The University of Tokyo, 5-1-5 Kashiwanoha, Kashiwa, Japan

**Abstract**

*We have carried out systematic fracture experiments in a single edge notch geometry over a range of stretch rates on dual crosslink hydrogels made from polyvinyl alcohol chains chemically crosslinked with glutaraldehyde and physically crosslinked with borate ions. If the energy release rate necessary for crack propagation was calculated conventionally, by using the work done to deform the sample to the critical value of stretch $\lambda_c$ where the crack propagates, we found that the fracture energy $\Gamma$ peaks around $\dot{\lambda} \sim 0.001 \, s^{-1}$ before decreasing sharply with increasing stretch rate, in contradiction with the measurements of crack velocity. Combining simulations and experimental observations, we propose therefore here a general method to separate the energy dissipated during loading before crack propagation, from that which is dissipated during crack propagation. For fast loading rates (with a characteristic strain rate only slightly lower than the inverse of the typical breaking time of physical bonds), this improved method to estimate a local energy release rate $g_{local}$ at the onset of crack propagation, gives a value of the local fracture energy $\Gamma_{local}$ which is constant, consistent with the constant value of the crack propagation velocity measured experimentally. Using this improved method we also obtain the very interesting result that the dual crosslink gels have a much higher value of fracture energy at low loading rates than at high loading rates, contrary to the situation in classical chemically crosslinked elastic networks.*

**Introduction**

Hydrogels have been used in pharmaceutical applications[1] for controlled drug release[2], in food science where their soft and elastic texture can be easily tuned to people's taste, and more recently, as model materials to mimic and replace living tissues[3,4]. Concurrently physicists have studied them to test rubber elasticity theories[5], equilibrium swelling[6] and as model poroelastic materials[7,8]. However, developing mechanically tough polymer gels and studying them has only become a popular research topic rather recently[9-11], stimulated by the discovery made by Gong et al.[12,13] that exceptionally tough gels could be made by introducing "sacrificial bonds" in the polymer networks. Their concept consists of a hard/brittle minority network embedded into a soft/deformable network forming two bicontinuous networks. When these double network gels are deformed, only the brittle network breaks and the microscopic bond breaking dissipates the strain energy without breaking the main network which is much more extensible, leading to remarkably high fracture energies $\Gamma \sim 1000$ J/m$^2$ [14-16]. A problem of the double network gels is that the broken bonds do not recover and so permanent damage remains after deformation[17,18]. To overcome this limitation, several research groups have recently synthesized doubly cross-linked tough gels with permanent and reversible crosslinks[19-22]. The reversible bonds break during deformation and serve as sacrificial bonds, but can also be reformed so that materials can recover their mechanical properties upon unloading. The mechanical toughness of the self-healing gels originates from the breaking and healing of these reversible crosslinks.

      While many articles report how to synthesize novel tough gels[23-27] and their basic mechanical properties in tension or compression, only a few focus on the physics of fracture of gels including transient bonds[20,28-30]. In particular, some studies report extraordinary toughening for gels that also display significant inelastic deformation[22,31]. How to account for this irreversible deformation in the evaluation of fracture energy is at the core of the present study.

The most important and difficult problem is how to relate the viscoelasticity resulting from the breaking/healing of the reversible crosslinks to the macroscopic fracture energy. In the case of completely elastic elastomers, the fracture energy to propagate a crack can be estimated by considering how much strain energy is needed to break the covalent bonds of polymer chains at the crack tip during the fracture process[32,33]. On the other hand, when both covalent and reversible crosslinks exist in polymer networks, even when the crack is not propagating some of the elastic strain energy that is imposed on the material is dissipated by the breakage of the reversible crosslinks during the loading stage, before the crack starts to propagate. Only the remaining strain energy can then be used to stretch and break polymer chains to create new interfaces in front of the crack tip during propagation. During crack growth, the region of energy dissipation is not necessarily confined to the crack tip and can actually occur throughout the entire sample. As is well known in viscoelastic fracture, the amount of dissipation is coupled to the fracture process even for a linear viscoelastic solid[34]. To separate the energy dissipated from the energy needed for fracture, modelling the nonlinear viscoelasticity of the self-healing gels is necessary.

In previous studies, we have synthesized a simple model dual crosslink gel with covalent and reversible crosslinks[35-37], and investigated experimentally and theoretically its time-dependent mechanical properties. These dual crosslink gels show a single relaxation time corresponding to the breaking of the reversible crosslinks, which considerably simplifies the physical modeling of the relationship between the bond breaking/healing and the macroscopic mechanical behaviors. Based on a theoretical model by Hui and Long[38], Long et al. proposed a multi-axial constitutive model to describe the nonlinear viscoelasticity of the dual crosslink gels by explicitly accounting for the breaking/healing kinetics of the reversible bonds[35]. By fitting stress relaxation data at a constant strain and stress-strain data at a constant strain rate with this model, the characteristic breaking/healing times of the breakable crosslinks can be estimated. From these characteristic times, the model can be used to predict stress-strain behavior for arbitrary strain histories including loading-hold-unloading cycles, showing almost perfect agreement between model predictions and experimental data[35]. Recently, we demonstrated that this constitutive model can also accurately capture the mechanical behavior of the dual cross-link gel samples under torsion[39]. The agreement between theory and experiments for both tension and torsion tests under different loading histories gives additional evidence that this model can be used to describe multi-axial loading histories, such as those in fracture experiments. This ability to make quantitative prediction is a great advantage of the dual crosslink gels system we used compared with other doubly cross-linked systems in order to understand more complex situations like fracture. For example, this model can be used to estimate the stored and dissipated energy during deformation for different strain histories.

In this work, we have carried out fracture experiments on pre-notched specimens of the dual crosslink gels and have analyzed the experimental data in light of the model developed by Long et al. The fracture tests give us the critical stretches when the crack starts to propagate and, from the time-dependence of the stress signal during fracture, we can also determine an average crack propagation velocity at various loading rates. We will demonstrate that the physical model describing loading-unloading curves can separate in certain situations the energy dissipated during loading by viscoelastic relaxation from the energy dissipated for bond breakage during crack propagation providing therefore a better estimate of the real elastic energy release rate needed to propagate the crack.

**Experiment**

**Materials**

The dual crosslink gels are made from polyvinyl alcohol (PVA) chains cross-linked by both covalent crosslinks and borate ions (Figure 1 (a)). The borate ions are mobile and create reversible crosslinks which detach from and reattach to PVA chains. Details of gel synthesis and chemistry of the reversible attachment are described elsewhere[36]. Briefly, the concentration of PVA (Aldrich, molecular weight = 89,000 - 98,000) in the dual crosslink gels was 12 %. The covalent crosslinks were introduced by adding glutaraldehyde (Aldrich) and the molar ratio between the chemical cross-linker and PVA monomers was 0.002 in the feed. The chemically cross-linked gels were soaked in borax/NaCl aqueous solutions (borax concentration: 1mM, NaCl concentration: 90 mM) to incorporate the borate ions in the PVA networks. For comparison, we prepared pure chemical gels with the same PVA and glutaraldehyde (chemical crosslinker) concentrations as those of the dual crosslink gels (Figure 1(b)). Comparing the purely chemically crosslinked gels with the dual crosslink gels reveals the effect of the reversible crosslinks on the fracture behavior.

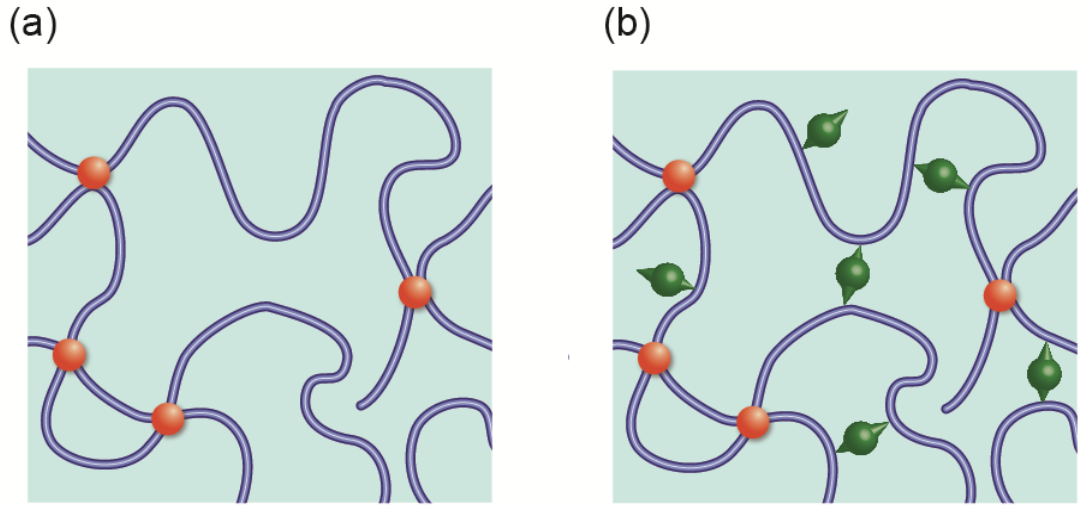

Figure 1. Schematic figure of (a) chemical gels and (b) dual crosslink gels having chemical (red) and physical (green) crosslinks. They have the same polymer concentration and chemical crosslinking density.

**Mechanical tests**

The tensile tests on un-notched and notched gel samples were performed on an Instron 5565 tensile tester with a 10 N load cell. Rectangular specimens of the dual crosslink and chemical gels are stretched until rupture at constant strain rates. Sample geometry of pre-notched specimens is shown in Figure 2. For the pure chemical gels, the sample thickness $h$, width $w$, initial length between clamps $l_0$, and notch length $c$ were 1.5 mm, 10 mm, 40 mm, and 2 mm, respectively. All the experiments on the chemical gels were conducted in air. In the case of the dual crosslink gels, the tensile tests were done in an oil bath in order to prevent drying and change strain rates over a wide range. The sample size of the dual crosslink gels was as follows: $h = 1.5$ mm (thickness), $w = 5$ mm (width), $l_0 = 20$ mm (length), and $c = 1$ mm (notch length).

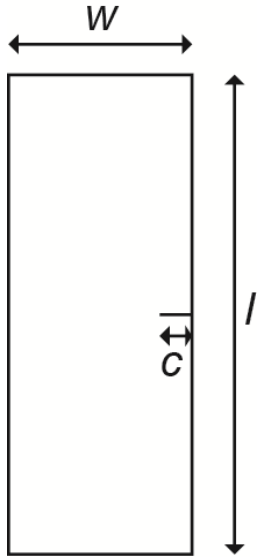

Figure 2. Schematic view of sample geometry for single edge notch tests.

**Results**

Figure 3 (a) and (b) show typical stress strain curves for the unnotched and notched gels, respectively. The tests were carried out at a fixed crosshead speed to impose a stretch rate $\dot{\lambda} = \dot{l}/l_0$ varying from 0.9 s$^{-1}$ to 0.0001 s$^{-1}$ for the dual crosslink gels and from 0.2 s$^{-1}$ to 0.01 s$^{-1}$ for the chemical gels. Here a dot denotes time derivative and $l$ is the gap between clamps. For the case of the chemical gels, there is little effect of the stretch rate on their mechanical response and fracture toughness as shown in Figure 3.

(a) un-notched samples

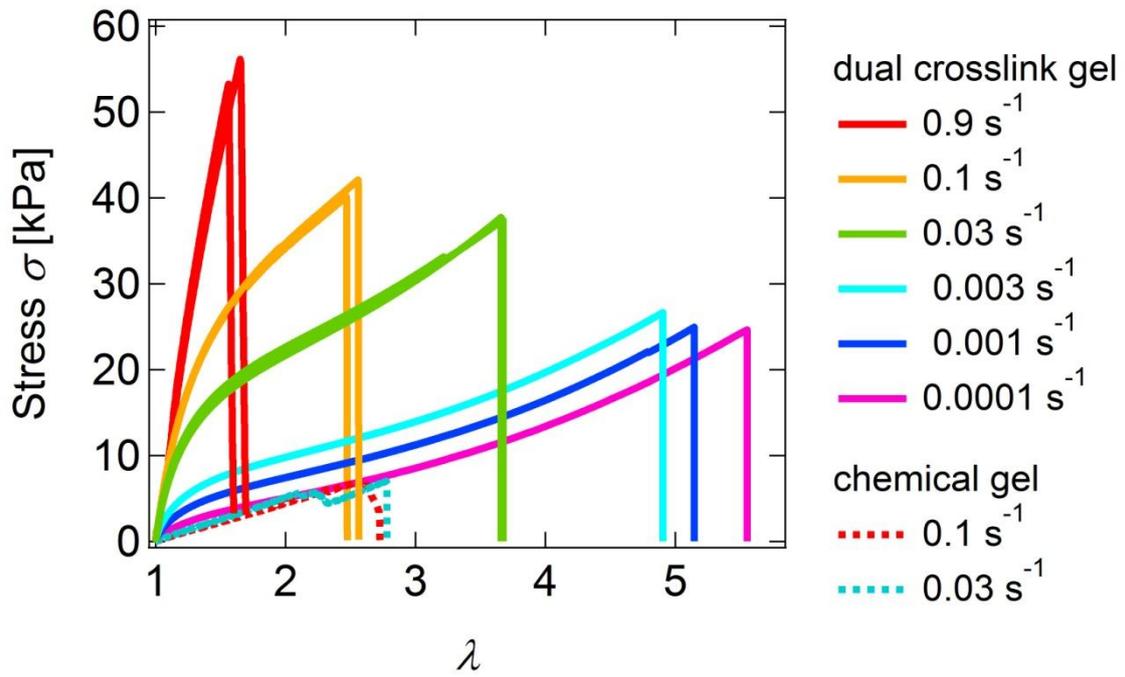

(a) un-notched samples

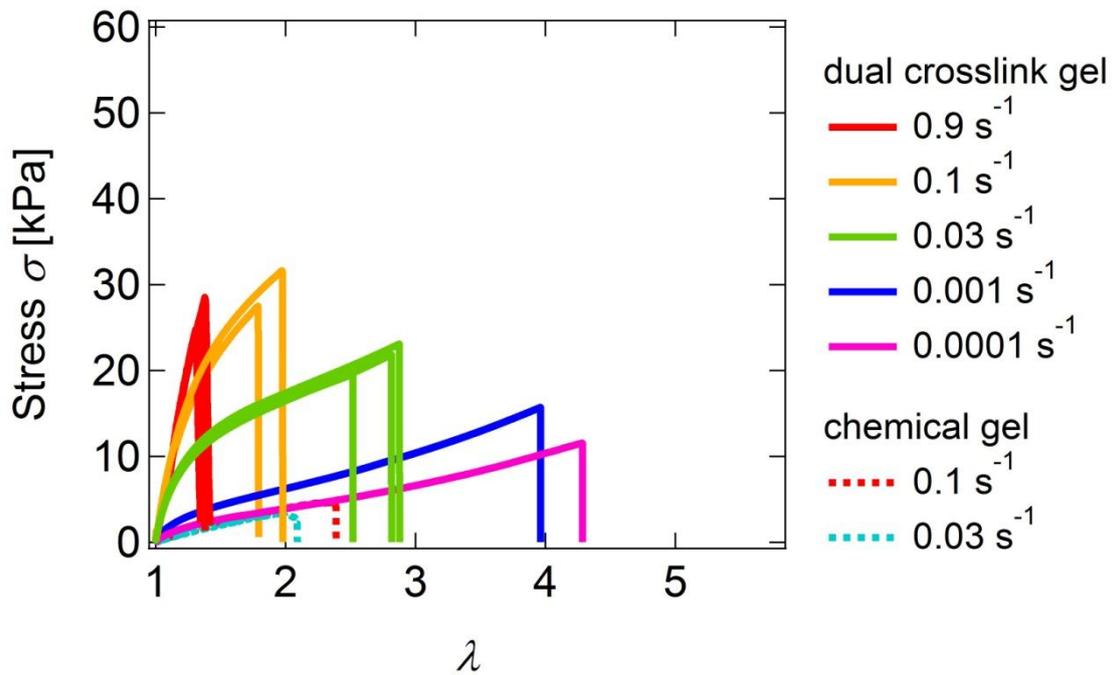

(b) notched samples

Figure 3. Stress versus stretch curves for (a) un-notched and (b) notched chemical gels and dual-crosslink gels at different stretch rates. Experiments were repeated two to three

times for all samples except the slowest velocity. Several curves are shown to illustrate reproducibility.

In comparison, the dual crosslink gels, show significant strain rate dependence due to the breaking and healing of the transient crosslinks. The critical stretch $\lambda_c$ at which fracture (crack propagation) starts is plotted against the stretch rate for the un-notched and notched dual crosslink gels in Figure 4. In both cases $\lambda_c$ increases with decreasing stretch rate and saturates at around 0.01 s$^{-1}$. $\lambda_c$ of the un-notched dual crosslink gels is about 30 % larger than that of the notched samples regardless of stretch rate.

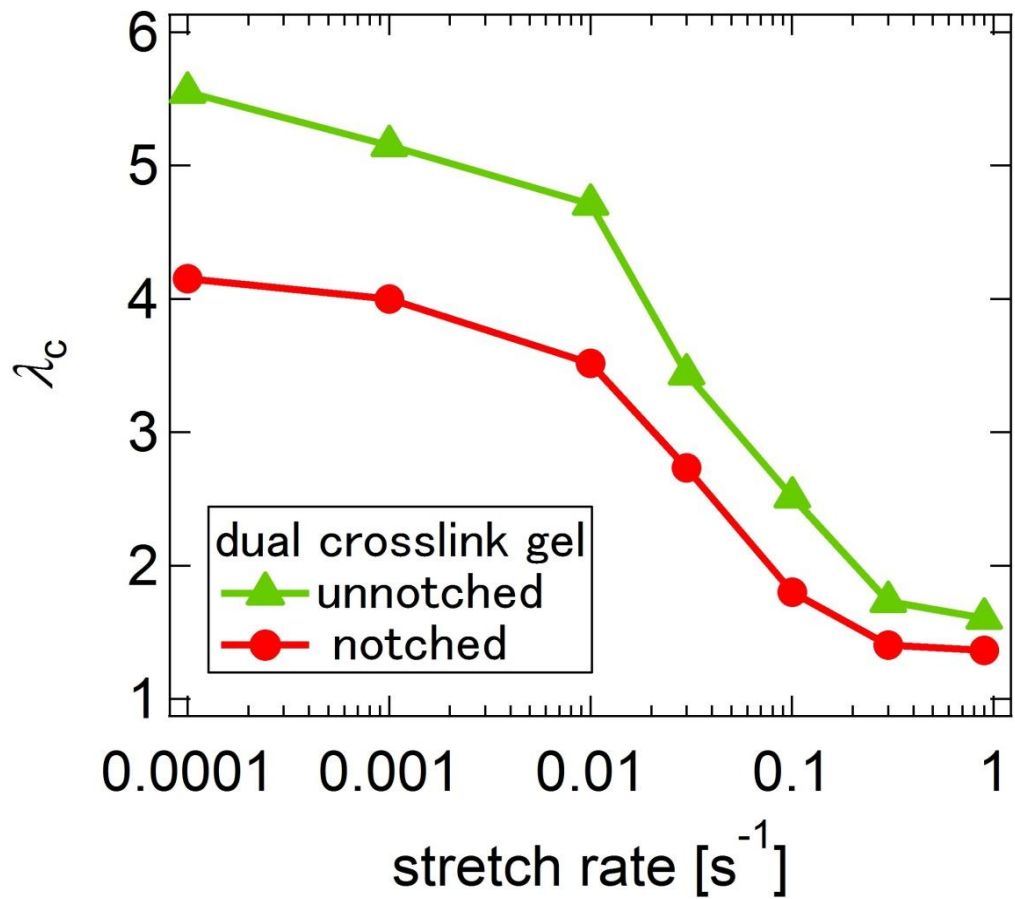

Figure 4. Stretch rate dependence of critical extension ratio $\lambda_C$ at which fracture starts for un-notched and notched dual crosslink gels.

It should be noted that at the lowest stretch rate, 0.0001 s$^{-1}$, the rate of breaking and reforming the reversible bonds is much faster than the loading rate, so that it is impossible for these bonds to accumulate any significant stretch before breaking and the amount of energy dissipation during loading is negligible. Thus, the stress vs. stretch curve before crack propagation is practically the same, as shown in Figure 3(b), for both chemical and dual cross-linked gels. Yet, the dual crosslink gels can be stretched up to $\lambda_c = 5$ (see Fig. 4) which is at least 2 times larger than $\lambda_c$ of the chemical gels (see Fig. 3b).

Figure 5 compares $\lambda_c$ between the notched dual crosslink and notched chemical gels. At higher strain rates, the dual crosslink gels rupture at lower stretch ratios than the chemical gels. Below 0.03 s$^{-1}$ $\lambda_c$ of the dual crosslink gels becomes larger than that of the chemical ones.

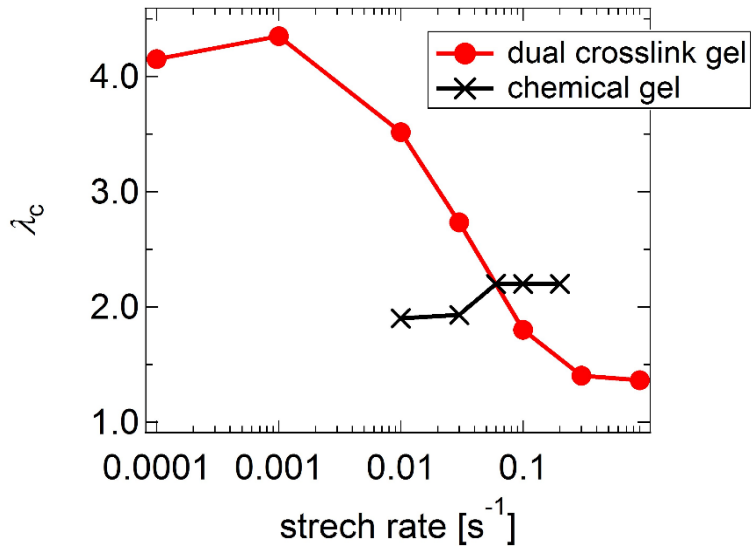

Figure 5. Stretch rate dependence of critical extension ratio $\lambda_C$ at which fracture starts for notched dual crosslink and notched chemical gels.

In order to compare quantitatively the fracture toughness of the chemical and dual crosslink gels, let us first estimate the fracture energy $\Gamma$ using an expression proposed by Greensmith[40] to calculate the energy release rate. Specifically, for a *small* crack of initial length $c$ in a purely elastic single edge crack specimen, the energy release rate $\mathcal{G}$ has to be proportional to $c$, since it is the only relevant length scale in the problem, thus

$$\mathcal{G} = 2\frac{3}{\sqrt{\lambda}}cW(\lambda), \qquad (1)$$

where $W(\lambda)$ is the stored energy density of an *un-notched* sample subjected to a simple uniaxial stretch $\lambda$.    Physically, equation (1) states that *all* the strain energy stored within a characteristic length of *c* is available for fracture.   Greensmith[40] has verified experimentally the validity of (1) for different elastomers loaded at moderate strains. The crack propagates for $\lambda = \lambda_c$ and for that condition we usually write $G = \Gamma$.    In the studies on elastomers $W(\lambda_c)$ has been estimated from the area under the nominal stress vs. stretch curve of the *un-notched* samples from $\lambda = 1$ to $\lambda = \lambda_c$:

$$W(\lambda_c) = \int_1^{\lambda_c} \sigma_{\text{un-notched}}\, d\lambda \qquad (2)$$

From eq. 1, 2 and Figure 3, the fracture energy $\Gamma$ for the chemical and dual crosslink gels is calculated as shown in Figure 6.

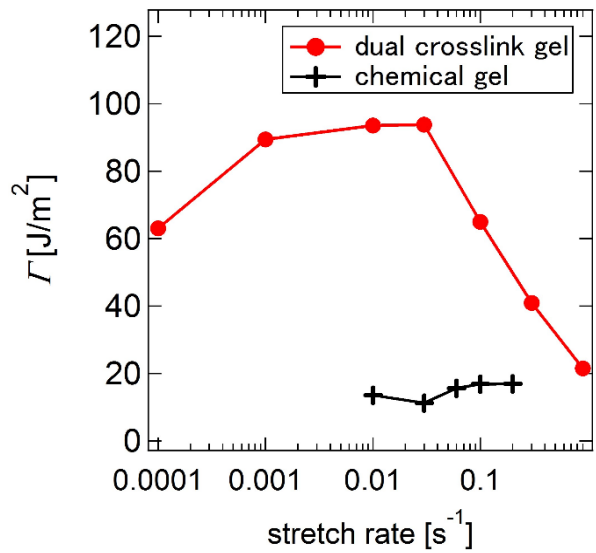

Figure 6. Plot of fracture energy $\Gamma$ calculated by eq 1 and 2 as a function of stretch rate.

Figure 6 shows $\Gamma$ of the chemical gels *remains constant* at about 20 J/m$^2$ while that of the dual crosslink gels depends *strongly on the stretch rate*.   However, it must be emphasized that eq 2 is really only valid for fully elastic materials although eq 1 and 2 , for lack of better options, have been applied to various kinds of polymeric materials including viscoelastic solids[20,22,41] . In the case of viscoelastic materials like the dual crosslink gels, a part of applied mechanical energy (work) is not stored elastically but dissipated during loading, so in general eq 2 overestimates $W(\lambda_c)$. .
    To proceed further, we attempt to separate the dissipated energy during loading

from the total applied work to estimate the total available energy for fracture. Let us look at how the crack propagates at high stretch rates. When a pre-notched gel is stretched, crack propagation starts almost instantaneously from $\lambda_c$ and the stress drops rapidly down to zero. Figure 7 (a) is a typical curve showing stress versus time for the tensile tests on the notched dual crosslink gels.

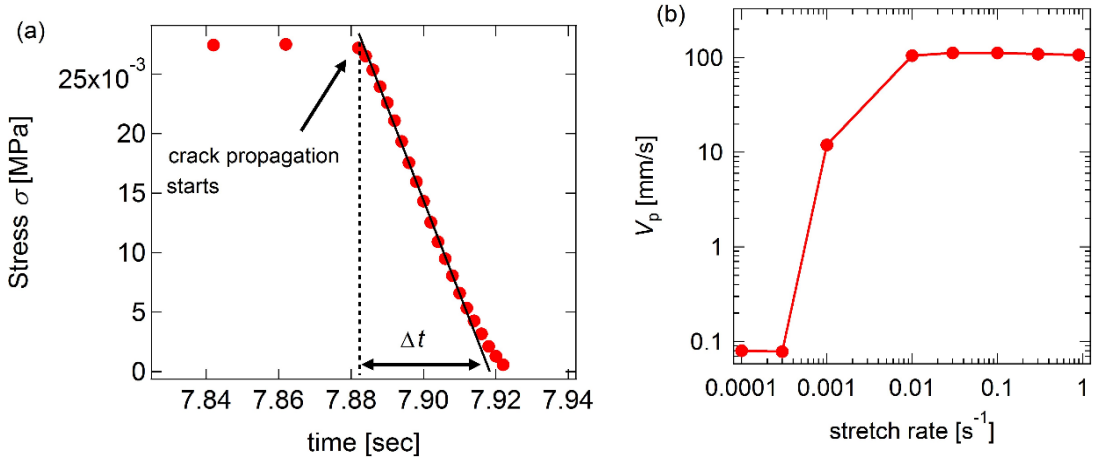

Figure 7 (a) Plot of stress against time for the single edge notch test of the dual crosslink gel at 0.1 s$^{-1}$, (b) stretch rate dependence of crack propagation velocity $V_p$ for the single edge notch tests on the dual crosslink gels.

From the initiation of the crack propagation, the nominal stress measured by the load cell decreases almost linearly with time. Using the time interval $\Delta t$ that the crack takes to pass through the specimen, we can determine the average velocity $V_p$ of the crack propagation:

$$V_P = \frac{w-c}{\Delta t}, \qquad (3)$$

where $w$ is the width of the specimen (Figure 2). As shown in Figure 7 (b), the crack propagation velocity $V_p$ of the dual crosslink gels stays constant at about 100 mm/s at stretch rates above 0.01 s-1, and slows down drastically at lower strain rates. Interestingly, in the range of strain rates from 0.01 s$^{-1}$ to 0.9 s$^{-1}$, despite the increase of the fracture energy $\Gamma$ shown in Figure 6, the crack propagation speed $V_p$ is independent of the stretch rate. This result is a further indication that $\Gamma$ calculated from eq 1 and 2 is incorrect and that the actual energy release rate during crack propagation should be constant - approximately independent of stretch rate for $\dot{\lambda} > 0.01$ s$^{-1}$.

In our experiments, the crack growth rate $V_p$ is a key factor to know how much energy is dissipated in the fracture process. Our hypothesis is that all the energy that is available for fracture is still contained in the same characteristic length $c$, except that $W$ in equation (1) is no longer given by (2) since it does not account for the fact that the energy available for fracture depends on the loading and unloading rates. In the following, we describe a procedure to determine an effective $W$ which allows us to compute the effective energy release rate. As discussed in the introduction, in an elastic solid, the loading and unloading curves are identical, irrespective of the loading and unloading rates, hence all the area under the stress versus stretch curve in (2) is available for fracture. To estimate the amount of energy available for fracture, we need to specify the unloading rate to compute the effective $W$. The unloading rate $\dot{\lambda}_U$ can be estimated as follows:

$$\dot{\lambda}_U = \frac{\lambda_c - 1}{\Delta t} = \frac{\lambda_c - 1}{w - c} V_p \ . \qquad (4)$$

The unloading rate $\dot{\lambda}_U$ for $\dot{\lambda} = 0.01$ to $0.9$ s$^{-1}$ varies from 66 to 9 s$^{-1}$; much higher than the inverse of the characteristic time for bond breaking and healing, that is, the physical bonds are fixed in their positions within the network during the fracture propagation. This is important, since it means that crack growth should not introduce significant changes in the configuration of the physical bonds ahead of the crack tip in these samples, giving credence to the use of an effective $W$ in which the loading rate is calculated using the nominal loading rate of the sample.

For the dual crosslink gels used in these tests, we have developed a quantitative theory to describe their non-linear viscoelastic mechanical response[35]. Using our model and the measured nominal loading rate and unloading rates given by (4), we can predict loading/unloading curves in an unnotched sample, and hence the amount of strain energy which is stored elastically and dissipated in the loading/unloading process. Figure 8 displays loading/unloading curves computed by the model for the loading rate $\dot{\lambda} = 0.9$ s$^{-1}$ (the unloading rate $\dot{\lambda}_U = 9.7$ s$^{-1}$) and the loading rate $\dot{\lambda} = 0.01$ s$^{-1}$ (the unloading rate $\dot{\lambda}_U = 66$ s$^{-1}$). The parameters in the model, such as breaking /healing times and elastic modulus, were obtained by fitting a relaxation experiment and a continuous loading experiment following the procedure described in[35].

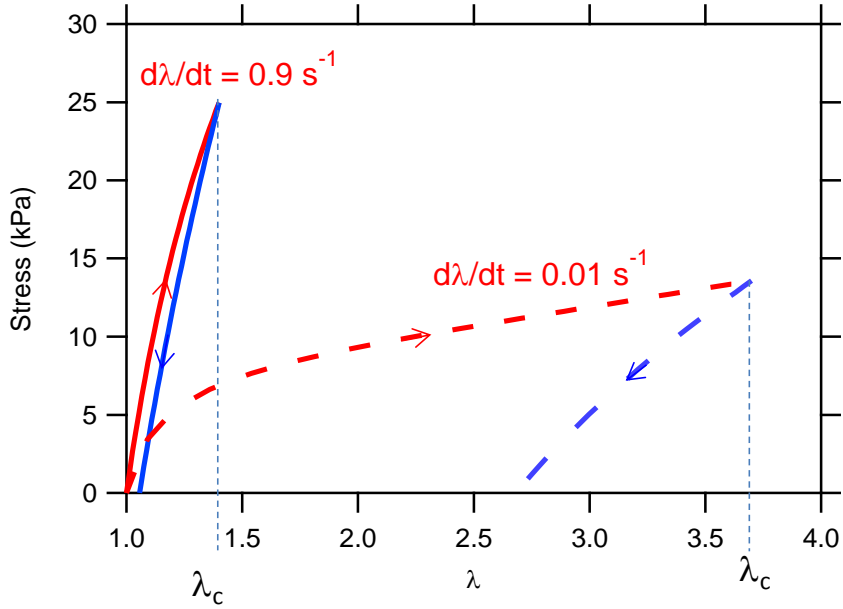

Figure 8 Calculated loading/unloading curves in uniaxial tension up to the critical stretch where the crack propagates in notched samples. The curves are computed with the model in [35] at two different stretch rates and corresponding unloading rates obtained from crack velocities. Note that the integral under the unloading blue curves is nearly identical while the integral under the red curves is not.

The area in the hysteresis loop corresponds to the dissipated strain energy while the area under the unloading curves is our estimate of the energy available for crack growth. Replacing $W(\lambda_c)$ in eq 1 by the area under the unloading curves gives the actual energy available for crack growth, a quantity we called the local energy release rate $g_{local}$. If our hypothesis is correct, then it should be equal to the actual fracture energy $\Gamma_{local}$ which is expected to be independent of the stretch rate.

| $\dot{\lambda}$ [s$^{-1}$] | $g_{local}$ [J/m$^2$] | $\lambda_c$ |
|---|---|---|
| 0.9 | 22.12 | 1.36 |
| 0.3 | 18.66 | 1.40 |
| 0.1 | 20.73 | 1.8 |
| 0.03 | 22.85 | 2.73 |
| 0.01 | 21.84 | 3.52 |
| 0.001 | 48.83 | 4.35 |
| 0.0003 | 39.09 | 3.6 |
| 0.0001 | 52.81 | 4.15 |

Table 1. Computed energy release rates for the dual crosslink gels.

Interestingly, as shown in Table 1 and Figure 9, $\Gamma_{local}$ calculated in this way, for $\dot{\lambda} = 0.01$ - 0.9 s$^{-1}$, is almost independent of the stretch rate. This is consistent with the fact that the crack propagation velocity $V_p$ is constant in this stretch rate regime. Surprisingly, even though the stress vs. stretch curves are quite different, the computed local fracture energy for $\dot{\lambda} = 0.9 – 0.01$ s$^{-1}$ is about 20 J/m$^2$, which is very close to the value of $\Gamma$ of the pure chemical gels.

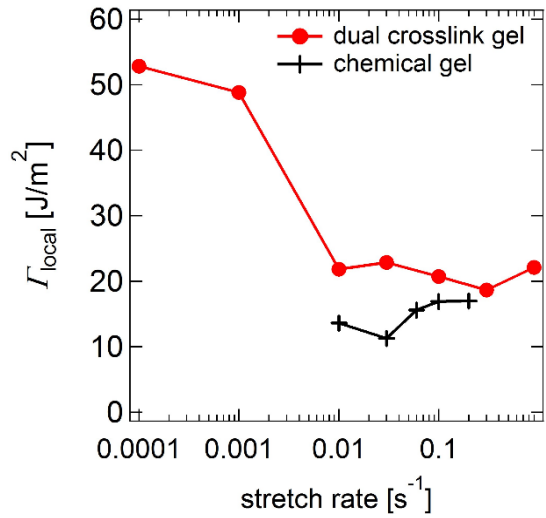

Figure 9 Plot of true fracture energy $\Gamma_{local}$ calculated by the model as a function of stretch rates.

At very low stretch rates, below 0.001 s-1, $\Gamma_{local}$ of the dual crosslink gels increases to around 50 J/m$^2$ and becomes closer to the classically measured $\Gamma$ (60 J/m$^2$ at 10$^{-4}$ s$^{-1}$ stretch rate from figure 6). In this regime the crack velocity is three orders of magnitude slower and certainly involves significant breakage and reformation of physical crosslinks not only during the loading but also near the crack tip *during the propagation of the crack*. At this slow stretch rate the material far from the tip is fully relaxed and behaves more like the chemical gel since all the original reversible bonds are broken, and the reformed bonds do not have time to accumulate any significant strains before they are broken again. Therefore, in this regime $\mathcal{G}$ (given by (1)) and $g_{local}$, calculated with the procedure described above, become increasingly closer to each other as $\dot{\lambda}$ decreases and so are $\Gamma$ and $\Gamma_{local}$.

**Discussion**

A curious fact is that the observed crack growth rate of our dual cross-link gels *decreases* with the fracture energy $\Gamma$. This result is markedly different from the case of traditional soft viscoelastic solids such as elastomers, which typically show a crack growth rate *increasing* with fracture energy. For simple hydrocarbon elastomers, values of $\Gamma$ typically range from about 100 J/m$^2$ to 100 kJ/m$^2$ and satisfy the well known empirical expression[42,43]:

$$\Gamma = \Gamma_0 \bigl(1 + \phi(a_T v)\bigr), \qquad (5)$$

where $\Gamma_0$ is the fracture energy at zero crack growth rate $v = 0$, $\phi$ is a monotonically increasing function of its argument with $\phi(0) = 0$ and $a_T$ is a temperature dependent shift factor which can be determined experimentally[44]. For elastomers, the minimum fracture energy $\Gamma_0$ is approached at very slow crack growth rate while the reverse holds for our dual cross-link gels, despite the fact that in both cases the materials are fully relaxed. This seemingly contradictory result can be explained by examining the local fracture process. In elastomers, the rate dependence of the fracture energy comes mainly from molecular friction near the crack tip, which increases as the crack growth rate increase or the temperature decreases. However in our gels molecular friction plays a minor role (water has a low viscosity) and the energy is dissipated by breaking bonds and releasing the strain energy of individual elastic strands between crosslinks.

At fast loading rates ($\dot{\lambda} > 0.01$ s$^{-1}$) energy is dissipated during the loading stage (accounting for the difference between $\Gamma$ and $\Gamma_{\text{local}}$) but during the fast fracture process one would expect both the physical and chemical bonds to be highly stretched near the crack tip. Since the areal density of bonds crossing the fracture plane is higher in the dual cross-link gel due to the presence of the physical bonds, one would expect a higher fracture energy. However, the presence of fully loaded physical bonds decreases the effective chain length between chemical cross-links, which, according to Lake and Thomas[45], has the effect of reducing the fracture energy. In our experiments, it seems that these two opposing effects cancel each other, resulting in a fracture energy $\Gamma_{\text{local}}$ close to that of the pure chemical gel.

At very low loading rates the bulk material behaves as the equivalent chemical gel and the physical bonds are invisible. However there must be a region *near the crack tip* where the physical bonds cannot be fully relaxed and must eventually share the load with the chemical bonds. In other words, our picture that healed physical bonds do not carry load cannot be true near the crack tip. This means that our constitutive model cannot be applied to material points near the crack tip. Indeed, our model assumes that the breaking and healing of physical bonds is independent of the strains carried by these bonds. While this assumption is supported by our tension and shear rheology experiments, it must be noted that these tests are carried out under small and moderate strains which is no longer the case near the crack tip. Very close to the crack tip, strain rates become very high and failure should resemble that of the double network gel of Gong[13], with the physical bonds acting as the breakable network and the chemical cross-links as the soft and extensible network; resulting in a much higher critical stretch ratio of $\lambda_c \approx 5$ at our slow loading rates.

As a final note we should discuss some of the limitations of our approach. For a rate-dependent material, the amount of energy dissipated at each material point depends on its stress history, which differs from point to point due to the stress gradient induced by the crack. As a result, the amount of energy actually available for fracture cannot be computed based on the loading and unloading history of a *remote point* from the crack tip. Our procedure works well however for fast crack growth rate since practically all the physical bonds in our sample are frozen in their current configuration during crack growth. The situation is expected to be different in the intermediate and slow crack growth regime, where material points surrounding the growing crack tip are undergoing different loading and unloading rates (those nearer to the crack tip is expected to load and unload much faster) and hence the configuration of physical bonds can vary significantly from one position to another in the crack tip region.

Furthermore we used a test geometry (single-edge notch) where the energy release rate is expected to increase with crack length (and yet we observe nearly a constant crack velocity during propagation). Experiments with pure shear specimens should be conducted to confirm that $\Gamma_{local}$ is indeed a material property independent of the geometry and this is the subject of ongoing work.

**Conclusion**

We show that the fracture energy of notched samples of dual crosslink gels made from

PVOH crosslinked chemically with a glutaraldehyde group and physically by borate ions, depends markedly on strain rate. If the critical energy release rate for crack propagation is calculated conventionally, by using the work done to stretch the sample to the value $\lambda_c$ where the crack propagates, we find that the fracture energy $\Gamma$ peaks around $\dot{\lambda} \sim 0.001$ before decreasing sharply with increasing stretch rate. On the other hand the crack propagation velocity is either very low: ($v \sim 100$ µm/s for $\dot{\lambda} \leq 2\times10^{-4}$ s$^{-1}$) or very high (100 mm/s for $\dot{\lambda} \geq 0.01$ s$^{-1}$) with a sharp transition between the two regimes.

We propose a method to separate the energy dissipated during loading *before crack propagation*, from that which is dissipated *during crack propagation*. For $\dot{\lambda} \geq 0.01$ s$^{-1+}$, this improved method gives a value of $\Gamma_{local}$ representative of crack propagation rather than crack initiation, which is constant, consistent with the constant value of the crack propagation velocity measured experimentally. Using this improved method we obtain two very interesting results: The dual crosslink gels have a much higher value of fracture energy at low loading rates than at high loading rates, contrary to the situation in classical chemically crosslinked elastic networks and the difference between $\Gamma$ and $\Gamma_{local}$ is the largest for intermediate stretch rates (0.01-0.03 s$^{-1}$), where most of the energy is dissipated during loading and not during crack propagation.

**References**

(1)  *Hydrogels in Medicine and Pharmacy*; Peppas, N. A., Ed.; CRC, 1986.
(2)  Peppas, N. A.; Hilt, J. Z.; Khademhosseini, A.; Langer, R. *Adv Mater* **2006**, *18*, 1345.
(3)  Peppas, N. A.; Langer, R. *Science* **1994**, *263*, 1715.
(4)  Slaughter, B. V.; Khurshid, S. S.; Fisher, O. Z.; Khademhosseini, A.; Peppas, N. A. *Adv Mater* **2009**, *21*, 3307.
(5)  Rubinstein, M.; Panyukov, S. *Macromolecules* **1997**, *30*, 8036.
(6)  Queslel, J. P.; Mark, J. E. *Adv Polym Sci* **1985**, *71*, 230.
(7)  Shibayama, M.; Tanaka, T. *Adv Polym Sci* **1993**, *109*, 1.
(8)  Shengqiang, C.; Zhigang, S. *EPL (Europhysics Letters)* **2012**, *97*, 34009.
(9)  Calvert, P. *Adv Mater* **2009**, *21*, 743.
(10) Zhao, X. *Soft Matter* **2014**, *10*, 672.
(11) Gong, J. P. *Soft Matter* **2010**, *6*, 2583.
(12) Tanaka, Y.; Gong, J. P.; Osada, Y. *Prog. Polym. Sci.* **2005**, *30*, 1.

(13) Gong, J. P.; Katsuyama, Y.; Kurokawa, T.; Osada, Y. *Adv Mater* **2003**, *15*, 1155.

(14) Tanaka, Y.; Kuwabara, R.; Na, Y. H.; Kurokawa, T.; Gong, J. P.; Osada, Y. *Journal of Physical Chemistry B* **2005**, *109*, 11559.

(15) Huang, T.; Xu, H. G.; Jiao, K. X.; Zhu, L. P.; Brown, H. R.; Wang, H. L. *Adv Mater* **2007**, *19*, 1622.

(16) Brown, H. R. *Macromolecules* **2007**, *40*, 3815.

(17) Webber, R. E.; Creton, C.; Brown, H. R.; Gong, J. P. *Macromolecules* **2007**, *40*, 2919.

(18) Wang, X.; Hong, W. *Soft Matter* **2011**, *7*, 8576.

(19) Tuncaboylu, D. C.; Argun, A.; Algi, M. P.; Okay, O. *Polymer* **2013**, *54*, 6381.

(20) Lin, W. C.; Fan, W.; Marcellan, A.; Hourdet, D.; Creton, C. *Macromolecules* **2010**, *43*, 2554.

(21) Li, J.; Illeperuma, W. R. K.; Suo, Z.; Vlassak, J. J. *ACS Macro Letters* **2014**, 520.

(22) Sun, J.-Y.; Zhao, X.; Illeperuma, W. R. K.; Chaudhuri, O.; Oh, K. H.; Mooney, D. J.; Vlassak, J. J.; Suo, Z. *Nature* **2012**, *489*, 133.

(23) Wang, X.; Wang, H.; Brown, H. R. *Soft Matter* **2011**, *7*, 211.

(24) Tuncaboylu, D. C.; Sari, M.; Oppermann, W.; Okay, O. *Macromolecules* **2011**, *44*, 4997.

(25) Kean, Z. S.; Hawk, J. L.; Lin, S.; Zhao, X.; Sijbesma, R. P.; Craig, S. L. *Adv Mater* **2014**, n/a.

(26) Carlsson, L.; Rose, S.; Hourdet, D.; Marcellan, A. *Soft Matter* **2010**, *6*, 3619.

(27) Haraguchi, K.; Uyama, K.; Tanimoto, H. *Macromol. Rapid Commun.* **2011**, *32*, 1253.

(28) Baumberger, T.; Caroli, C.; Martina, D. *Eur. Phys. J. E* **2006**, *21*, 81.

(29) Seitz, M. E.; Martina, D.; Baumberger, T.; Krishnan, V. R.; Hui, C.-Y.; Shull, K. R. *Soft Matter* **2009**, *5*, 447.

(30) Baumberger, T.; Ronsin, O. *The European Physical Journal E* **2010**, *31*, 51.

(31) Li, J.; Suo, Z.; Vlassak, J. J. *Journal of Materials Chemistry B* **2014**, *2*, 6708.

(32) Rivlin, R. S.; Thomas, A. G. *Journal of Polymer Science* **1953**, *10*, 291.

(33) Cristiano, A.; Marcellan, A.; Keestra, B. J.; Steeman, P.; Creton, C. *J Polym Sci Polym Phys* **2011**, *49*, 355.

(34) Schapery, R. A. *International Journal of Fracture* **1975**, *11*, 141.

(35) Long, R.; Mayumi, K.; Creton, C.; Narita, T.; Hui, C.-Y. *Macromolecules* **2014**, *47*, 7243.

(36) Narita, T.; Mayumi, K.; Ducouret, G.; Hebraud, P. *Macromolecules* **2013**, *46*, 4174.

(37) Mayumi, K.; Marcellan, A.; Ducouret, G.; Creton, C.; Narita, T. *ACS Macro Letters* **2013**, *2*, 1065.

(38) Hui, C.-Y.; Long, R. *Soft Matter* **2012**, *8*, 8209.

(39) Long, R.; Mayumi, K.; Creton, C.; Narita, T.; Hui, C.-Y. *Journal of Rheology (1978-present)* **2015**, *59*, 643.

(40) Greensmith, H. W. *Journal of Applied Polymer Science* **1963**, *7*, 993.

(41) Zhang, H.; Chen, Y. J.; Lin, Y. J.; Fang, X. L.; Xu, Y. Z.; Ruan, Y. H.; Weng, W. G. *Macromolecules* **2014**, *47*, 6783.

(42) Gent, A. N.; Schultz, J. *Journal of Adhesion* **1972**, *3*, 281.

(43) Maugis, D.; Barquins, M. *Journal of Physics D: Applied Physics* **1978**, *11*, 1989.

(44) Gent, A. N. *Langmuir* **1996**, *12*, 4492.

(45) Lake, G. J.; Thomas, A. G. *Proceedings of the Royal Society of London, series A: Mathematical and Physical Sciences* **1967**, *A300*, 108.